\documentstyle[aps,twocolumn,prb,epsf]{revtex} 

\begin{document}

\draft
\twocolumn[    
\hsize\textwidth\columnwidth\hsize\csname @twocolumnfalse\endcsname    

\title{Gap ratio in anharmonic charge-density-wave systems
}
\author{J. K. Freericks$^{1}$ and Veljko Zlati\'c$^2$}
\address{$^1$Department of Physics, Georgetown University, 
  Washington, DC 20057-0995, U.S.A.\\
$^2$Institute of Physics, Zagreb, Croatia}
\date{\today}
\maketitle

\widetext
\begin{abstract}
Many experimental systems exist that possess charge-density-wave order
in their ground state.  While this order should be able to be described 
with models similar to those used for superconductivity, nearly all systems
have a ratio of the charge-density-wave order parameter to the transition
temperature that is too high for conventional theories.  Recent work 
explained how this can happen in harmonic systems, but when the lattice 
distortion gets large, anharmonic effects must play an increasingly
important role.  Here we study the gap ratio for anharmonic charge-density
wave systems to see whether the low-temperature properties possess
universality as was seen previously in the transition temperature and to
see whether the explanation for the large gap ratios survives for anharmonic
systems as well.
\end{abstract}

]      

\narrowtext
\section{Introduction}

The appearance of stripe ordering in materials related to the 
high-temperature superconducting oxides\cite{htsc_cdw} and of charge
and orbital ordering in the colossal magnetoresistance 
materials\cite{colossal_cdw}, has provided a renewed interest in the 
physics that drives charge-density-wave order.  This phenomenon has been
seen in a wide variety of materials ranging from quasi-one-dimensional
systems (where Peierls distortion physics is important) such as the
organic conductors\cite{organic_cdw}, to the di- and 
trichalcogenides\cite{selenides_cdw}
like 2H-TaSe$_2$ or NbSe$_3$, so-called A15 materials\cite{a15_cdw} such
as V$_3$Si, blue bronzes\cite{bbronze_cdw} like KMoO$_3$, cubic 
oxides\cite{bkbo_cdw} like Ba$_{1-x}$K$_x$BiO$_3$, and Verwey transition
materials\cite{verwey_cdw} such as Fe$_3$O$_4$.

There has been much theoretical work on this problem as well.  The 
dynamical mean field theory was employed to solve for the charge-density-wave
phase in the ordered state\cite{italians} and the puzzle of the large gap
ratio was resolved for harmonic systems\cite{millis_cdwratio}. The resolution
involves properly accounting for phonon renormalizations, for nonvanishing
effects from vertex corrections, and from effects due to nonconstant electronic
density of states (since conventional Migdal-Elisahberg approaches
cannot produce gap ratios $2\Delta/T_c$ larger than about 8).  The anharmonic
problem was also examined with dynamical mean-field 
theory\cite{freericks_universal} and it was found that
the transition temperature satisfied a scaling law with the wavefunction
renormalization parameter for a wide range of parameter space (including
systems that were not Fermi liquids at low temperatures).

Here we concentrate on the issue of whether or not the gap ratio can remain
large when anharmonic effects are present.  The scaling law for the
transition temperature shows that anharmonicity does not have a dramatic effect
on $T_c$, but as the system is cooled down to $T=0$, the lattice distortion 
becomes larger and larger generating the full charge-density-wave gap.  
Naively, we would expect anharmonicity to reduce the lattice distortion
(relative to a harmonic system) because the higher powers in the phonon 
potential do not allow the phonon coordinate to move as far away from the
origin. Hence, one expects that anharmonicity will generically reduce the
gap ratio, since $T_c$ will be unaffected, but $\Delta$ will be reduced
relative to the results of a harmonic system with the same value of
the wavefunction renormalization parameter.
We need to verify whether this effect occurs and determine how large it
can be to see whether 
one can still explain the large gap ratios of charge-density-wave systems
in the presence of anharmonic potentials. 

In Section II we introduce the model and the techniques used to solve
for the gap ratio.  Section III contains our results and discussion, and
Section IV contains our conclusions.

\section{Formalism}

We will be investigating the static anharmonic Holstein model, whose 
Hamiltonian is\cite{holstein,freericks_universal}
\begin{eqnarray}
H&=&-\sum_{i,j,\sigma}t_{ij}c^{\dagger}_{i\sigma}c_{j\sigma}+\sum_i
(g\bar x_i-\bar \mu)(n_{i\uparrow}+n_{i\downarrow})
\cr
&+&\frac{1}{2}\bar\kappa
\sum_i\bar x_i^2+\alpha_{an}\sum_i\bar x_i^4.
\label{holstein_ham1}
\end{eqnarray}
Standard notations are used here: $c^{\dagger}_{i\sigma}$ ($c_{i\sigma}$)
creates (destroys) an electron at lattice site $i$ with spin $\sigma$, 
$n_{i\sigma}=c^{\dagger}_{i\sigma}c_{i\sigma}$ is the electron number
operator, $\bar\mu$ is the chemical potential, and $\bar x_i$ is the phonon
coordinate at lattice site $i$. We examine the static (or classical) phonon
case here, so there is no kinetic energy of the phonon.  The hopping of the
electrons is restricted to nearest neighbors on a hybercubic lattice in
$d$-dimensions.  We take the limit $d\rightarrow\infty$ and scale $t_{ij}=
t^*/(2\sqrt{d})$ in order to have nontrivial results\cite{metzner_vollhardt}.
The rescaled hopping integral $t^*$ determines our energy scale ($t^*=1$).
The bare density of states becomes a Gaussian $\exp(-\epsilon^2)/\sqrt\pi$,
with $\epsilon$ the band energy.  The local phonon is taken as a classical
variable, so it only has a spring constant $\kappa$ associated with it. The
anharmonic potential is chosen to be of the simplest form, a quartic term
with a strength $\alpha_{an}$. The deformation potential (or electron-phonon
interaction strength) is denoted by $g$ and measures an energy per unit length.

It is useful to shift the phonon coordinate, in order to see explicitly the
particle-hole symmetry present in
the harmonic model.  We shift $\bar x_i\rightarrow
x_i+x^\prime$ with $g\langle n\rangle
+\bar\kappa x^{\prime}+3\alpha_{an}x^{\prime 3}=0$ (with $\langle n\rangle$ the 
average total electron filling), to transform
the Hamiltonian into
\begin{eqnarray}
H&=&-\sum_{i,j,\sigma}t_{ij}c^{\dagger}_{i\sigma}c_{j\sigma}+\sum_i
(g x_i- \mu)(n_{i\uparrow}+n_{i\downarrow}-\langle n\rangle)\cr
&+& \frac{1}{2}\kappa \sum_i x_i^2+\beta_{an}\sum_ix_i^3+
\alpha_{an}\sum_i x_i^4,
\label{holstein_ham2}
\end{eqnarray}
with $\kappa=\bar\kappa+12\alpha_{an}x^{\prime 2}$, $\beta_{an}=
4\alpha_{an}x^{\prime}$, and $\mu=\bar\mu-gx^{\prime}$.  It is the presence
of the cubic term, when $\alpha_{an}\ne 0$ that removes the particle
hole symmetry\cite{hirsch}
from the problem when $\mu=0$ and $\langle n\rangle=1$,
since the Hamiltonian is no longer unchanged under the transformation 
$n_{i\sigma}\rightarrow 1-n_{i\sigma}$ and $x_i\rightarrow -x_i$.  Note that
the system still will possess charge-density-wave order at half filling and
weak coupling (for all small $g>0$) because the band structure is still nested
at half filling, even though the Hamiltonian is not particle-hole symmetric.

Our calculations are performed using standard techniques of dynamical
mean-field theory\cite{italians,millis_cdwratio}: we iterate a series of 
equations to self-consistency that involve (i) determining the local Green's
function from the self energy (by integrating over the noninteracting density
of states), (ii) extracting the effective medium (by removing the self energy
from the local Green's function),
(iii) calculating the probability distribution of the phonon coordinate 
(by solving the atomic path integral in a time-dependent field),
(iv) extracting the electronic self energy (after determining the local
Green's function from an integral over the 
phonon coordinate distribution).  This procedure is standard (even for the
case of the ordered phase), and we do
not describe any further details here. Our calculations are performed
at half filling $\langle n\rangle =1 $.  In the harmonic case, we have
$\mu=0$, but the anharmonic problem must have the chemical potential
adjusted as a function of temperature to yield the right filling.

When the system is in an ordered charge-density-wave state,
there are two probability distributions for the phonon coordinate---one for the
$A$ sublattice $w_A(x)$ and one for the $B$ sublattice $w_B(x)$.  The order
parameter for the charge-density-wave phase is defined to be
\begin{equation}
\Delta(T)=g\int dx [w_A(x)-w_B(x)]x,
\label{delta_def}
\end{equation}
which measures the average difference in the phonon coordinate between the $A$
and $B$ sublattices multiplied by the deformation potential.

\begin{figure}[htb]
\vspace{5mm}
\epsfxsize=3.0in
\centerline{\epsffile{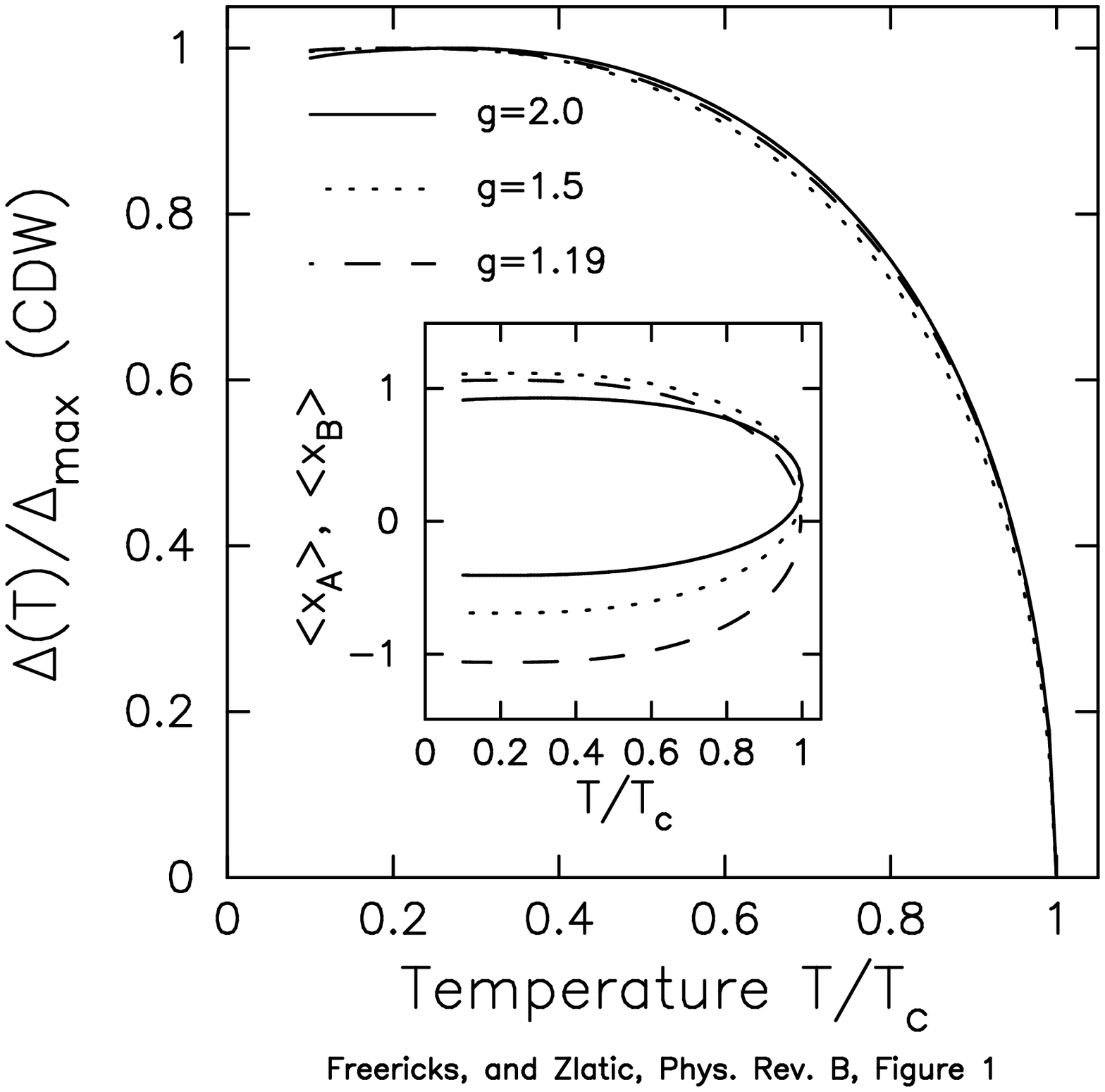}}
\caption{Charge-density-wave order parameter as a function of temperature
for three cases of anharmonicity (all chosen to be at the maximal
$T_c$) (i) $g=2$, $\alpha_{an}=0.16$, $T_c=0.1277$ (solid line); (ii)
$g=1.5$, $\alpha_{an}=0.03$, $T_c=0.1326$ (dashed line); and (iii)
$g=1.19$, $\alpha_{an}=0$, $T_c=0.1340$ (dotted line).  The figure
is renormalized to plot $\Delta(T)/\Delta_{max}$ and $T/T_c$.  Inset
is a plot of the average phonon coordinate on the $A$ and $B$ sublattices.
Note that all the curves are symmetric, but their average value moves
away from zero for the anharmonic cases.}
\end{figure}

In Figure 1 we show a plot of this charge-density-wave order parameter as
a function of temperature.  The plots include three cases, all at the maximum 
value of $T_c$ for a given value of $g$ and arbitrary $\alpha_{an}$.  The
parameters chosen are $g=2$, $\alpha_{an}=0.16$ which has $T_c=0.1277$;
$g=1.5$, $\alpha_{an}=0.03$ which has $T_c=0.1326$; and $g=1.19$, 
$\alpha_{an}=0$ which has $T_c=0.1340$.  The curves are normalized by the
maximal gap value calculated at $T\approx T_c/8$ which is very close to the 
$T=0$ value,
and by $T_c$.  Note that rather than approaching 1, there is a small decrease
in $\Delta(T)/\Delta_{max}$ at the lowest temperatures.  This is an accuracy
issue with our calculations (which are performed with a fixed number of 
5000 Matsubara frequencies for the energy cutoff).  We estimate that our error 
in the $T=0$
gap is no larger than about 1\%. These curves have the correct generic behavior
that we expect for mean-field systems: the order parameter increases like
$\sqrt{T_c-T}$ away from the transition temperature and then rapidly saturates.
We see no significant difference in the shape of these curves for different
values of the anharmonicity.  Inset into this figure is a plot of
$\langle x_A\rangle$ upper curves  and $\langle x_B\rangle$ lower curves.
Note how the anharmonic systems are both shifted upwards and have narrower
spreads in the average phonon coordinate, but that the curves are symmetric
between the $A$ and $B$ sublattices, so the particle-hole asymmetry only affects
the midline of the phonon coordinate, not its distortion in the ordered phase!

In summary, we can accurately estimate the value of the $T=0$ gap by performing 
calculations at $T=T_c/10$ and the shape of the order parameter (as a
function of $T$) is not too
strongly dependent on the strength of the anharmonicity.  In addition, we
find that the main effect of the anharmonic interaction is to change the 
average of the phonon coordinate values on each sublattice $(\langle x_A\rangle
+\langle x_B\rangle)/2$ and shrink the magnitude of the distortion.  This does
not mean that the gap is reduced by as much, though, because systems that
share the same approximate value of $T_c$ will have different values of
both $g$ and $\alpha_{an}$.  The reduction in the distortion of the
phonon coordinate can be compensated by a correspondingly larger value of $g$.

\section{Results and Discussion}

We now present results for the charge-density-wave gap as a function of
the anharmonicity.  The most reasonable way to present these results
is to plot the gap versus a measure of the anharmonic potential energy
in equilibrium.  We can determine what the equilibrium phonon coordinate
$\bar x =x^*$ is for the atomic problem [determined by 
Eq.~(\ref{holstein_ham1}) with $t_{ij}=0$ and $\langle n\rangle=1$]
and then plot results versus
$\alpha_{an}x^{*4}$. These results are summarized in Fig.~2.  Note that
this measure of
the anharmonic potential energy is double valued when $g$ is large, since it 
approaches
0 in both the small and large $\alpha_{an}$ limits. The general shapes
of these curves are quite similar, but the scales change with the
coupling strength and the stronger coupled cases show more curvature (and
eventually a double-valuedness).

\begin{figure}[htb]
\epsfxsize=3.0in
\centerline{\epsffile{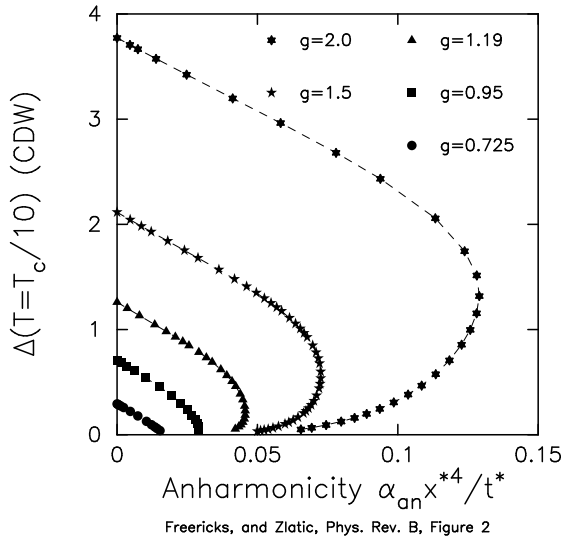}}
\caption{Charge-density-wave gap calculated at $T_c/10$ plotted versus the
anharmonic ``energy'' $\alpha_{an}x^{*4}$ determined from the average value
of the phonon coordinate for the atomic problem.  Note how the different
curves ($g=$ 0.725, 0.95, 1.19, 1.5, and 2.0)
have the same general shape, which become double valued for the
strongest coupling cases shown here.}
\end{figure}

We can examine the static Holstein model in both a 
strong-coupling\cite{freericks_strong} and a weak-coupling\cite{millis_weak}
limit.  In the strong coupling limit, the transition temperature
approaches zero as $1/(2g[x_0-x_2])$, while the zero temperature gap becomes 
large as $g(x_0-x_2)/2$ and hence the ratio can diverge
[$2\Delta/(k_BT_c)=2g^2(x_0-x_2)^2$]. Here $x_i$ satisfies
\begin{equation}
g(i-1)+\kappa x_i + 3 \beta_{an}x_i^2+4\alpha_{an}x_i^3=0.
\label{xdef}
\end{equation}
This ratio becomes infinite as the coupling strength
increases, hence there is no limit to the magnitude of the gap ratio.
Note that this result is also true for the superconducting order, when
one includes the effect of a nonconstant density of states, since the
strong-coupled $T_c$ will approach zero there as well. This result comes 
entirely from the fact that the phonon-coordinate distortion grows linearly
with $g$ in strong coupling, but the transition temperature decreases as
$1/g^2$ due to the strong-coupling physics.

We can also investigate the possibility of universal behavior in strong
coupling.  Since the system consists of empty sites and preformed pairs, the
self energy has a low-energy pole and takes the form $\Sigma(\omega)=\alpha
/(\omega+i\delta)+O(\omega^0)$ for small $|\omega|$ with $\alpha>0$.
Plugging this form into the self-consistent equations for the Green's function
yields $\alpha=-1/2+g^2|x_0||x_2|$ (which is larger than zero for $g$
large enough).  Defining the wavefunction renormalization parameter
by a scaling along the imaginary axis\cite{freericks_universal}
\begin{equation}
Z=1-\left \{ \frac{3}{2}\frac{{\rm Im}\Sigma(i\omega_0)}{\omega_0}-
\frac{1}{2}\frac{{\rm Im}\Sigma(i\omega_1)}{\omega_1}\right \},
\label{zdef}
\end{equation}
with $i\omega_j=i\pi T(2j+1)$, yields
\begin{equation}
Z=1-\frac{13}{18}\frac{1}{\pi^2T^2}+\frac{13}{9}\frac{g^2|x_0||x_2|}{\pi^2T^2}.
\label{zstrong}
\end{equation}
Since $Z$ is a function of $x_0$ and $x_2$ and $2\Delta/(k_BT_c)$ is a function
of $x_0-x_2$, we don't expect universal behavior at extremely strong coupling
for the anharmonic case.  But, in the harmonic case we have $|x_0|=|x_2|$,
so we expect deviations from universality only when $|x_0|/|x_2|$ deviates
far from unity.

The weak-coupling
limit is much more complicated.  There are many approaches that can
be taken\cite{millis_weak,freericks_weak}, but none produce good agreement
with the transition temperature over a wide range of coupling strengths
(but the zero-temperature gap is approximated well).
Here we will concentrate on just two different strategies (for
the harmonic case only): (i) the renormalized
phonon method\cite{millis_weak} where a certain class of vertex correction
terms can be neglected from the analysis, but one needs to work with
renormalized phonons (renormalized by the electron-hole bubble diagrams)
and (ii) a similar approximation\cite{freericks_weak} that employs the 
identical set of diagrams, but does  not renormalize the phonons.

\begin{figure}[htb]
\epsfxsize=3.0in
\centerline{\epsffile{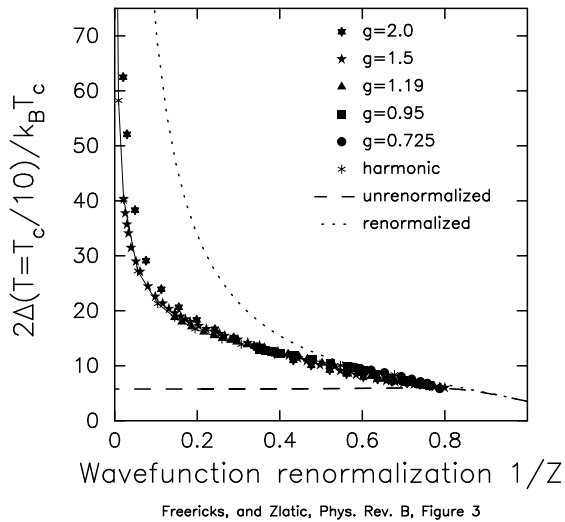}}
\caption{Ratio of twice the charge-density-wave gap (calculated at $T_c/10$) to
the charge-density-wave transition temperature $T_c$  plotted versus the
wavefunction renormalization parameter. The harmonic limit is connected
by a thin solid line, while the renormalized (dotted) and unrenormalized
(dashed) weak coupling approximations are also included.  Note how the
anharmonic systems generically lie close to the harmonic results for 
a wide range of parameters.  Deviations occur for large $g$ and for intermediate
values of the net coupling strength, but generally are no larger than about
20\% (except at very strong coupling where the gap ratio is {\it enhanced}).  }
\end{figure} 

In Fig.~3 we show the results for the gap ratio in the harmonic case,
five different anharmonic cases, and the two approximation schemes
described above.  In order to check for universal behavior, we plot
the results as a function of the wavefunction renormalization 
parameter\cite{freericks_universal} $Z$ [which is evaluated as in 
Eq.~(\ref{zdef})]. One can see some striking behavior 
in this plot.  First, we find that the gap ratio can become as large as one
would like as the coupling strength gets bigger and bigger. Furthermore,
the gap ratio rises very rapidly above the weak-coupling limit of 3.52,
so even $Z$ factors of 1.25 (which would correspond to very weak 
electron-phonon coupling) would have a gap ratio larger than 6.
We find that the calculated results lie in between the two different
weak coupling schemes indicating that phonon renormalization is important,
but the summation of the bubble diagrams renormalizes the phonons too
strongly in the general case. Finally, we notice that except for the case
of very strong coupling ($g=2$), the gap ratio is modified by at most about 20\%
from the harmonic case. This finding is most surprising, but can be explained
in relatively simple terms.  As the anharmonic potential increases, the system
becomes weaker and weaker coupled, because the anharmonic potential prevents the
phonon coordinate from deviating far from the origin.  In the limit
of very weak coupling, the transition is dominated by a nesting instability
and can be described within a BCS-like format.  Hence all results must
agree in this limit (if one properly identifies the net
strength of the attractive 
interaction).  Furthermore, when the system has small anharmonicity,
the anharmonic effects can be treated perturbatively, and the system remains
close to the harmonic limit there as well.  Since the curves are pinned to be
close to the harmonic limit for small anharmonicity and for large 
anharmonicity, we find that they generically do not stray far from the
harmonic curve in the intermediate regime.  As the coupling strength
increases so this intermediate regime becomes larger, the deviations can also
become significant, as we see for the $g=2.0$ case.  But we find it surprising
that there is such a wide range of parameter space where the results for the
anharmonic system remain so close to the harmonic system!

\section{Conclusions}

Contrary to the simple arguments about anharmonicity, we find that generically
anharmonic phonons are described well by an equivalent harmonic limit, even in
the ordered phase, where one would expect the effects of the anharmonicity
to be felt more strongly.  Hence, we find that the analysis given for
the large gap ratio in the harmonic electron-phonon problem remains essentially
unchanged as anharmonicity is introduced except in the limit of extremely
strong electron-phonon coupling (which may be so unphysically large that is
is not attained in any real material).  The reason why this holds is essentially
a continuity argument: for small and large anharmonicity, the system must be
close to the harmonic results---hence it remains close for intermediate
values as well. 

These results imply, once again\cite{freericks_universal}, that the
quasiharmonic approximation should be quite accurate in real materials, since
the anharmonic system can be mapped onto an effective harmonic system that
shares both the single-particle and two-particle properties. Deviations are
expected only in the strong coupling and strong anharmonic limit.  The
deviation of the gap ratio from the BCS value can be understood from the
harmonic analysis\cite{millis_cdwratio}.  We don't expect these results to be 
changed much for small phonon frequencies, because we know that $T_c$ follows 
a universal
form for low frequency, and we expect the gap value will not change much 
if the phonon frequency is small.

\section*{Acknowledgments}
We would like to acknowledge stimulating discussions with Andy Millis.
This work was supported by the National Science Foundation
under grant DMR-9973225.


\end{document}